\documentclass[10pt,journal,compsoc]{IEEEtran}

\ifCLASSOPTIONcompsoc
  \usepackage[nocompress]{cite}
\else
  \usepackage{cite}
\fi

\usepackage{graphicx}
\usepackage{amsmath,amssymb}
\usepackage{booktabs}
\usepackage{listings}
\usepackage{multirow}
\usepackage{float}
\usepackage{fontspec}

\usepackage[table]{xcolor}
\usepackage[most]{tcolorbox}
\tcbuselibrary{skins}
\usepackage{enumitem}
\usepackage{url}
\usepackage{xurl}
\usepackage{stfloats}
\usepackage{fontawesome5}
\usepackage{microtype}
\usepackage{tikz,pgfplots}
\pgfplotsset{compat=1.18}

\usepackage{pifont}

\definecolor{ieeeblue}{RGB}{0,73,123}
\definecolor{softgray}{RGB}{245,246,248}
\definecolor{modelblue}{RGB}{230,240,255}
\definecolor{processgreen}{RGB}{240,250,230}
\definecolor{platformyellow}{RGB}{255,250,220}
\definecolor{orggray}{RGB}{245,245,245}
\definecolor{envmint}{RGB}{235,255,240}

\newtcolorbox{ScenarioBox}[1][]{%
  enhanced,
  colback=softgray,
  colframe=ieeeblue,
  boxrule=0.5pt,
  arc=2mm,
  left=2mm,right=2mm,top=1.2mm,bottom=1.2mm,
  drop shadow,
  borderline west={2.2pt}{0pt}{ieeeblue},
  title=\textbf{#1},
  fonttitle=\normalsize\sffamily,
}

\newtcolorbox{rqanswerbox}[1]{%
  enhanced,
  breakable,
  colback=white,
  boxrule=0pt,
  left=4mm,
  borderline west={1pt}{0pt}{black!70},
}

\usepackage[breaklinks=true]{hyperref}
\usepackage{breakcites}

\lstdefinestyle{code}{
  basicstyle=\ttfamily\footnotesize,
  breaklines=true,
  frame=single
}

\begin{document}

\title{Vibe Coding in Practice: Flow, Technical Debt, and Guidelines for Sustainable Use}

\author{%
Muhammad~Waseem,
Aakash~Ahmad,
Kai-Kristian~Kemell,
Jussi~Rasku,
Sami~Lahti,
Kalle~M\"akel\"a,
Pekka~Abrahamsson%
\IEEEcompsocitemizethanks{
    \IEEEcompsocthanksitem Muhammad Waseem, Kai-Kristian Kemell, Jussi Rasku, and Pekka Abrahamsson are with Gpt-Lab Tampere University, Finland.\\
    Email: \{muhammad.waseem, kai-kristian.kemell, jussi.rasku, pekka.abrahamsson\}@tuni.fi
    \IEEEcompsocthanksitem Aakash Ahmad is with the University of Derby, United Kingdom.\\
    Email: a.abbasi@derby.ac.uk
    \IEEEcompsocthanksitem Sami Lahti is with Koivu Solutions, Finland.\\
    Email: sami.lahti@koivusolutions.com
    \IEEEcompsocthanksitem Kalle M\"akel\"a is with Aimbition, Finland.\\
    Email: kalle.makela@aimbition.com
}
\thanks{Manuscript prepared for IEEE Software.}
}

\IEEEtitleabstractindextext{%
\begin{abstract}
Vibe Coding (VC) is a form of software development assisted by generative AI, in which developers describe the intended functionality or logic via natural language prompts, and the AI system generates the corresponding source code. VC can be used for rapid prototyping or developing the Minimum Viable Products (MVPs); however, it may introduce several risks throughout the software development life cycle. Based on our experience from several internally developed MVPs and a review of recent industry reports, this article analyzes the flow–debt tradeoffs associated with VC. The flow–debt trade-off arises when the seamless code generation occurs, leading to the accumulation of technical debt through architectural inconsistencies, security vulnerabilities, and increased maintenance overhead. These issues originate from process-level weaknesses, biases in model training and data, a lack of explicit design rationale, and a tendency to prioritize quick code generation over human-driven iterative development. Based on our experiences, we identify and explain how current model, platform, and hardware limitations contribute to these issues, and propose countermeasures to address them – informing research and practice towards more sustainable VC approaches.
\end{abstract}

}

\maketitle
\IEEEdisplaynontitleabstractindextext

\section{Introduction}
\label{sec:introduction}
\IEEEPARstart{V}{ibe coding}, a term previously unknown to the software engineering community, was brought into the spotlight by Andrej Karpathy\footnote{\url{https://www.klover.ai/andrej-karpathy-vibe-coding/}} in early 2025. VC refers to an intuitive, flow-driven programming supported by generative AI. VC can be seen as the next stage in the evolution of AI-assisted programming, moving beyond autocomplete and code completion toward a conversational-driven software development. Enthusiasts view VC as a way to improve productivity by accelerating prototyping and lowering barriers for individuals with limited programming experience \cite{ge2025survey}.

We use the term VC to describe multi-turn, intent-driven development in which natural-language prompts drive the creation, regeneration, and restructuring of full software systems. VC is fundamentally different from IDE auto-completion, GitHub Copilot-style in-line suggestions, or occasional prompts for small code fragments. It moves control from isolated line-level assistance to conversational generation, automatic scaffolding, and rapid restructuring of end-to-end systems.

With today’s more capable models, developers can express intent in natural language and incrementally build working systems within minutes. What was once a speculative idea has become routine in prototypes, hackathons, and early-stage commercial development. Recent surveys indicate that most practitioners perceive AI-assisted tools as improving productivity and workflow efficiency \cite{stackOverflowDevSurvey2024}. This growth resembles earlier trends in low-code and no-code platforms, where rapid prototyping enables broader participation and faster exploration of design alternatives.

 \textbf{Potential and concerns of VC}: VC increasingly influences the entire Software Development Life Cycle (SDLC). By 2025, platforms such as Replit, Cursor, Bolt, Kiro, and Lovable began supporting end-to-end system development, illustrating rapid adoption across open-source and commercial ecosystems \cite{replit2025growth}. While these tools enable conversational, AI-driven development at scale, recent practitioner reports highlight recurring concerns related to reliability, security, and long-term maintainability \cite{palmer2025vulnerabilities, noble2025fragility, futurecode2025risks}.


Although individual developers and enterprises are rapidly adopting VC tools, major concerns remain regarding the reliability, security, and maintainability of generated systems. Recent analyses show severe vulnerabilities, missing documentation, and architectural fragility in many AI-generated applications \cite{palmer2025vulnerabilities, noble2025fragility, futurecode2025risks}. For example, Palmer et al.\ identified misconfigurations in more than 170 applications created with the Lovable platform, exposing sensitive data \cite{palmer2025vulnerabilities}. Practitioners similarly report that while VC accelerates prototyping, it often produces code that is difficult to scale or maintain in production settings \cite{mahendra2025vibecoding, futurecode2025risks}.

\begin{figure}[h]
    \centering
    \includegraphics[width=1\linewidth]{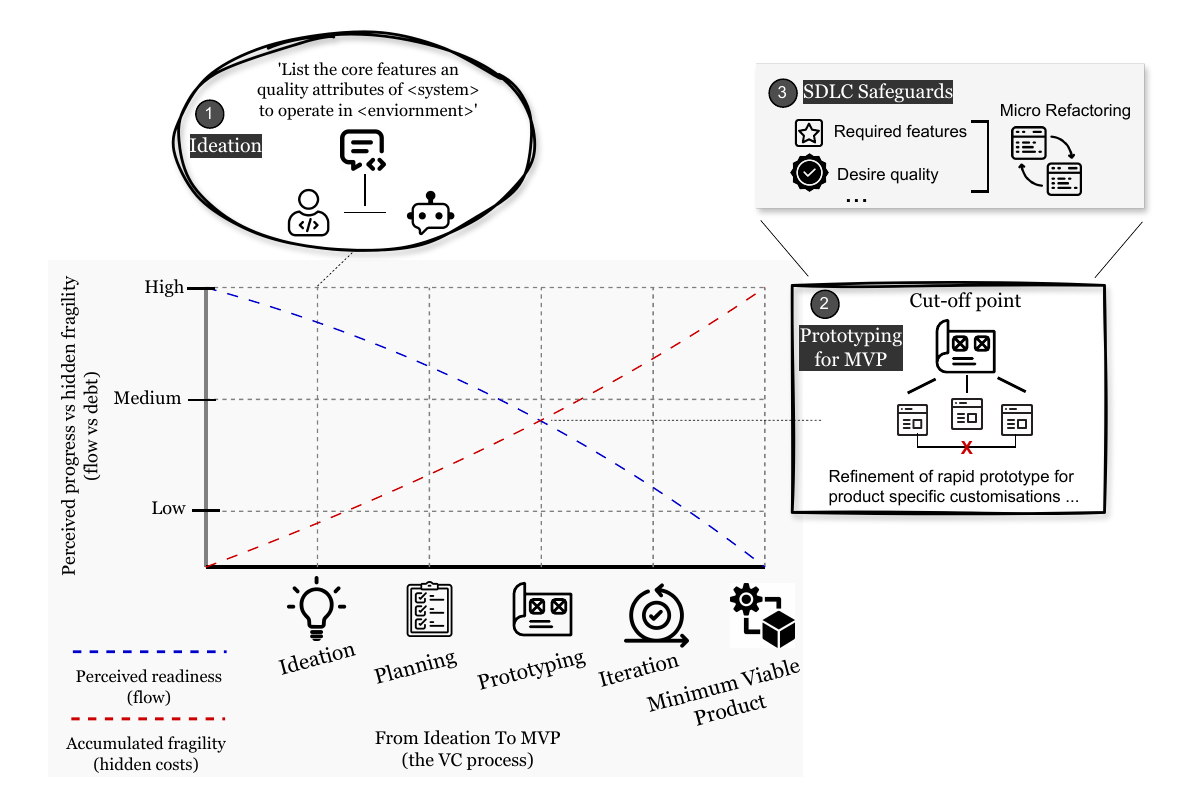}
    \caption{Flow–debt tradeoff over a project timeline (illustrative concept; not empirical or to scale).}
    \label{fig:flowdebt}
\end{figure}

This creates a \textbf{flow–debt tradeoff}: VC accelerates short-term development but risks the accumulation of technical debt that undermines long-term sustainability. Understanding this tradeoff is essential for practitioners, researchers, and tool builders alike. Figure~\ref{fig:flowdebt} illustrates how rapid progress during ideation and prototyping is often followed by rising technical debt as systems move toward MVP maturity, highlighting the need for SDLC guardrails such as code inspections and targeted micro-refactoring.


\begin{figure*}
    \centering
\includegraphics[width=0.67\linewidth]{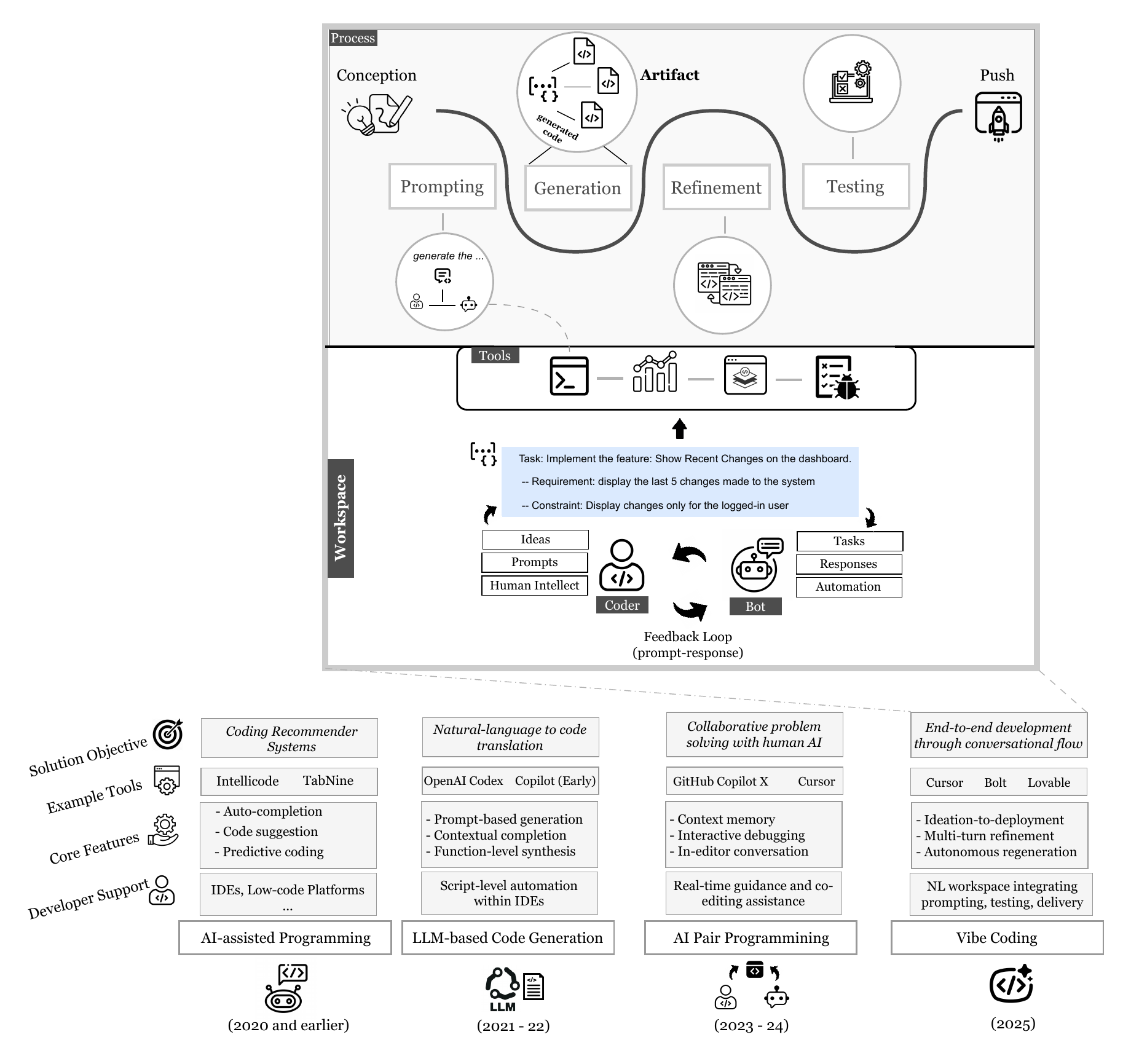}
    \caption{End-to-end process of VC: transforming ideas into artifacts through human–AI collaboration. }
    \label{fig:End-to-end}
\end{figure*}

\textbf{Contribution and Implications:} 
In this paper, we (i) conceptualize the flow–debt trade-off across the VC lifecycle, (ii) illustrate these mechanisms using experience-based case material, and (iii) synthesize practical guidance into a cross-layer matrix for sustainable VC covering model, process, platform, organizational, and environmental levels.

\subsection*{GenAI-Assisted Development Experiences} The observations in this article draw on the authors’ combined academic and industrial work with GenAI-assisted development. VC tools are used in practice for prototyping, refactoring, analysis, and building experimental and production systems, and our industry collaborators contribute insights from real-world product engineering and organizational adoption. Although the evidence is qualitative, it provides a coherent, practice-grounded basis for understanding the development behaviors, technical challenges, and emerging debt mechanisms described in this paper.

\section{Development Flow}
This study is primarily based on our direct experience as researchers and practitioners of vibe coding tools. We, along with several other members of our teams, frequently use vibe coding assistants such as Cursor AI to develop MVPs. Over the past few months, we have shared these experiences informally, in one-to-one discussions, casual conversations over coffee, and even during seminars and conferences where AI enthusiasts debated the promises and pitfalls of vibe coding. These discussion identified a recurring patterns. As illustrated in Figure~\ref{fig:flowdebt}, our own experiences echoed a pattern about early progress felt smooth and productive, but fragility accumulated beneath the surface. 
One of the co-authors of this study, who has extensive experience as a solution architect and developer, stated:

\begin{quote}
{\faQuoteLeft}  \textit{In our earlier development practices, communicating with business stakeholders relied heavily on methods such as Specification by Example and Behavior-Driven Development (BDD), where the scope of work is expressed through structured pseudocode and executable examples. These approaches constrain natural-language descriptions to be non-open-ended, reducing ambiguity and giving VC tools clearer boundaries for interpreting intent. As a result, early VC work shifts from purely requirement-driven to a more scope-driven approach, where the assistant can reason about the difference between the current system and the desired change. As VC platforms evolve, dialogue-based intelligent systems may increasingly support this pattern by capturing intent in a structured manner before generation begins, improving both clarity and downstream consistency} {\faQuoteRight}
\end{quote}


However, we also noted that when we started with a generic natural-language description, the assistant often went beyond our stated requirements, generating detailed plans and suggesting additional “fancy” features. Many of these proposals looked polished and attractive, giving the impression that a production-ready application was within reach. The generated MVPs created strong excitement: within a short time, we felt we had something close to Version 1, complete with appealing designs and functionality that seemed ready to demo.

To further illustrate this perceived momentum in practice, we shared a scenario from our own work, titled \textit{Fast Starts, Harder Finishes} (Figure~\ref{fig:refactoring-iteration-overload}). This scenario summarizes the early stages of developing a refactoring and code-analysis platform using a VC tool. It demonstrates how VC tools accelerate ideation and prototype creation through rapid prompt–response cycles, producing visually complete and functional modules within minutes. The perceived values observed during these early experiences are summarized as follows: 

\begin{itemize}
    \item \textbf{Ideation:} VC sped up idea generation by turning natural-language prompts into working code, often within minutes. Because the results arrived so quickly, it became easy to explore and refine ideas iteratively. For instance, \textit{when considering a ‘refactor preview’ feature, we asked, ‘How might developers visualize refactoring before applying it?’} We got back several suggestions about possible representations and interaction modes. This is really helpful for the team to refine the conceptual direction before development. 
    \item \textbf{Iteration:} This fast idea-to-implementation phase led to a more iterative workflow. Developers could quickly adjust prompts, regenerate improved code versions, and validate changes through automatic testing of the generated code. This cycle also provides immediate feedback for rapid refinement and maintaining project momentum. 
    \item \textbf{Human-AI collaborative development}: VC tools facilitated component integration and code reuse by making previously generated snippets or modules easy to reproduce and combine. This interaction between human developers and AI-generated components produced more uniform structures across the project and minimized the time teams spent reconfiguring or rebuilding elements in each iteration. In practice, developers could quickly assemble functional prototypes without handling low-level details, while the AI handled repetitive regeneration tasks. This steady reuse strengthened architectural coherence as the system evolved and created a smoother workflow between human judgment and AI assistance.
    \item\textbf{Prompt-based productivity}: Natural language prompts raise the level of abstraction in development work so that participants can contribute even without strong programming skills. This approach brings more people into the development process while maintaining meaningful output. In parallel, prompt-driven generation reduces repetitive coding and documentation tasks, giving experienced developers more time to focus on complex reasoning, architecture decisions, and overall workflow coordination.
    \item\textbf{Rapid prototyping and experimentation}: VC supported quick prototyping across different tools and environments, making it easy to start new projects or migrate existing ones with minimal overhead. This responsiveness strengthened both the flexibility and the perceived scalability of generative workflows. After defining the idea of a “\textit{refactor preview}”, for instance, a simple prompt such as “create an interface that shows original and refactored code side by side” produced a working visualization within minutes. This initial prototype provided a concrete way to evaluate the clarity and value of the concept before committing to detailed design or implementation.
\end{itemize}
Taken together, these layers of perceived value, ranging from ideation and iteration to integration, delegation, and experimentation, capture the holistic flow experience of vibe coding. This foundation of rapid creativity and inclusivity also reveals an emerging tension between speed and sustainability, which will be explored in the following section.

\begin{figure*}[t]
\centering
\begin{ScenarioBox}[Fast Starts, Harder Finishes - VC Scenario for Refactoring and Code Analysis Platform]
\small
\textbf{MVP Scenario:} We started building a refactoring and code analysis platform (https://github.com/wasimsse/RefactAI) as an early prototype to see how VC would work in practice. Our goal was to create a working MVP with key features, including a dashboard, code metrics visualization, and a refactoring preview interface.

\textbf{Rapid Development (the Flow):} In the start, progress was smooth. The prompts we provided to the system quickly yielded polished components. The dashboard, analysis view, and refactor interface came together easily, and each part worked well on its own. This stage demonstrated how quickly VC can generate components, with instant feedback and strong momentum. The prototype seemed to be demo-ready, and we did weeks of work in minutes.

\textbf{Technical Issues in Development (the Debt):} Problems started when we attempted to refine the system. When the analysis results did not display correctly, asking the VC assistant to fix them led to significant changes, including rewriting state management, adding new API endpoints, implementing additional error handlers, and creating debugging panels. Each time, the codebase grew by hundreds of lines, but the main issue persisted. Debugging became more challenging as the number of modules and the amount of overlapping logic increased. Ultimately, we identified the root cause by manually reviewing the code, which revealed a simple mismatch between the backend and frontend APIs. In VC, it is easy to add features quickly, but integrating everything adds complexity.

\end{ScenarioBox}
\label{fig:refactoring-iteration-overload}
\end{figure*}

\subsubsection*{When Vibe Coding is Appropriate}
The experiences described above indicate that vibe coding can be highly suitable in contexts where rapid exploration and low coordination cost are more important than long-term stability. In particular, VC aligns well with internal tools, early-stage ideation, UI mockups, proofs of concept, and experimental feature development. These settings benefit from the speed of natural language interaction, the ability to generate alternatives quickly, and the low overhead required to adjust direction in response to stakeholder feedback.

VC is also effective when teams need to explore several design options before settling on a stable direction. Its strengths in ideation, regeneration, and visual refinement help teams evaluate options without committing to fully engineered solutions. For organizations experimenting with new ideas or evaluating feasibility, VC can provide rapid insight and reduce friction in early development. This perspective sets the stage for understanding where the same properties that create value in early exploration can also introduce risks when applying VC to production-grade or safety-sensitive systems.

\section{Flow to Debt}
Vibe coding rapidly transforms ideas into executable artifacts through prompts and LLM assistance. Yet, this flow often comes with hidden costs. As systems evolve, the same mechanisms that turn rapid ideas into executable code introduce inconsistencies, ambiguities, and integration issues that accumulate into technical debt and other problems. This section examines how such debt emerges across different phases. Based on our experience, we summarize the most common symptoms.

\subsection{Debt: Requirements Ambiguity}

When initiating application development via VC, the process typically starts with describing the application through high-level prompts. While these prompts yield early requirements for the system, they are frequently inconsistent and contain multiple forms of ambiguity. This ambiguity primarily arises from stakeholders’ choice of wording, which the model may interpret differently from their intended meaning~\cite{ronanki2023requirements, vasudevan2025role}. Based on our observations, during application development the generated requirements may initially appear well structured but often include unnecessary features, omit important non-functional aspects, or introduce inconsistencies between requirements and design. These omissions occur most frequently for Non-Functional Requirements (NFRs) such as security, performance, reliability, and evolvability.

VC–assisting tools (e.g., Cursor AI\footnote{Mentioned as a representative platform; similar behaviours have been observed in other tools.}) can produce requirements that look polished and detailed, along with step-by-step implementation plans. When handled carefully, these outputs can be useful, yet they also pose a risk. As prototypes evolve into larger systems, the generated requirements may overwhelm teams with excessive detail. This raises a concern about the overall project and requirement management. Specifically, we face the challenge of collecting NFRs or obtaining vague NFRs, which are rarely captured as \textit{ASRs}, the subset of requirements that drive major design decisions. When ASRs are missing or vague, LLMs under the VC tools develop only based on visible requirements, and the missing or vague ASRs remain ignored until we define them clearly and in a structured way for every class, file, or module. In line with our observations, recent studies report that GenAI assistants focus primarily on functional aspects but provide weaker support for qualities such as security, performance, and maintainability~\cite{mey2024aiassistants}. Without safeguards, this ambiguity at the requirements stage has a ripple effect on downstream software development phases, which in turn accumulates into technical debt. One co-author of this study concluded his experience with VC-based requirements and proposed the solution:

\begin{quote}
{\faQuoteLeft} \textit{One of the challenges is that vibe coding can blur the line between requirements, pseudocode, and technical design. When AI generates new requirements or changes, they must be checked against existing regression tests and Behaviour-Driven Development (BDD) scenarios. The team needs to update these scenarios when the context changes so that the design stays consistent with the expected behaviour of the system.} {\faQuoteRight[solid]}
\end{quote}


Another practical approach is to treat NFRs as first-class citizens and explicitly mark ASRs. A lightweight template based on quality-attribute scenarios (stimulus, source, environment, affected artifact, response, and measurable response) can be embedded in prompts and linked to acceptance criteria, tests, and CI budgets (e.g., latency budgets, error budgets, or security policies). This approach creates a clear trace from ASRs to architectural decisions and regression checks.
\textit{For instance, prompts can be enhanced to ask: “Before generating code, identify potential ASRs and express them as quality-attribute scenarios. For each ASR, specify Stimulus (e.g., user request), Source (e.g., external client), Environment (e.g., production system), Artifact (e.g., the \texttt{/search} endpoint), Expected Response (e.g., average response time), and Measurable Criterion (e.g., P50 latency under 200,ms at 300 RPS).”}

\begin{tcolorbox}[
  enhanced,
  breakable,
  colback=white,
  boxrule=0pt,
  left=2mm,
  borderline west={1pt}{0pt}{black!80}
]
\textbf{Takeaway:} 
VC currently lacks systematic requirements engineering practices, leaving open how generated requirements can be made reliable, traceable, and verifiable. Prioritizing NFRs by identifying ASRs early, capturing them as quality-attribute scenarios with measurable fit criteria, and linking them to architecture and tests can help ensure that LLM-generated outputs align with long-term quality goals.
\end{tcolorbox}

\subsection{Debt: Architectural Inconsistency}
In VC, architecture is rarely established systematically, as development often moves directly from prompts to code. This leads to inconsistencies when modules are regenerated or extended, resulting in a patchwork structure rather than a coherent design~\cite{maes2025-gotchas}. Reviews confirm that while GenAI is increasingly used to bridge requirements-to-architecture and architecture-to-code, systematic evaluation and validation of architectural decisions remain weak, with recurring issues such as hallucinations and limited design reasoning~\cite{esposito2025-mlr}. 

We observed how VC increased these challenges. In our experience, when a microservices-based system was progressively developed through prompts, the resulting services frequently introduced integration issues that took many hours to resolve. For instance, in our experience, one microservice regenerated its authentication mechanism from JSON Web Tokens to session cookies, causing authorization failures (HTTP 401) across dependent services. In another case, regenerated API contracts used inconsistent field names, which broke frontend integration tests. Small inconsistencies in one service often cascaded across the system, requiring deep architectural and domain expertise to fix. The inconsistency issue shows the importance of established architectural practices for VC-based development. We also observed that several other architectural styles and principles, for example, \textit{layered separation of concerns} (to avoid mixing user interface, business, and data logic), \textit{architectural modeling and documentation} (for traceability and communication), \textit{quality attribute trade-off analysis} (to capture scalability, security, and performance constraints), and \textit{governance of architectural decisions} (to prevent drift across iterations) are often skipped when systems are generated directly from prompts. 
Based on extensive practical experience, one author of this study highlighted architectural styles that can help address this issue:
\begin{quote}
{\faQuoteLeft[solid]} \textit{We found that Domain-Driven Design (DDD) and Hexagonal Architecture (also known as Ports and Adapters Architecture) work particularly well with VC. These styles help keep responsibilities clear and support Test-Driven Development (TDD), Acceptance Test-Driven Development (ATDD), and BDD flows. They also make it easier to guide the assistant at the right level of abstraction, moving from high-level acceptance scenarios to Application Programming Interface (API) interfaces. This helps control the agent’s context window and reduces architectural drift.} {\faQuoteRight[solid]}
\end{quote}
Based on this observation, these challenges can be reduced by applying structured architectural practices such as DDD, Hexagonal Architecture, layered separation of concerns, architectural modeling, quality-attribute analysis, and clear governance to maintain stable boundaries during VC-based development.

\begin{tcolorbox}[
  enhanced,
  breakable,
  colback=white,
  boxrule=0pt,
  left=2mm,
  borderline west={1pt}{0pt}{black!80}
]
\textbf{Takeaway:}Architectural practices such as separation of concerns, documentation, quality attribute analysis, and decision governance remain essential. Structured approaches like DDD and Hexagonal Architecture can help maintain consistency by giving VC tools clearer boundaries and testable interfaces. Without these practices, inconsistencies that are manageable in MVPs can turn into serious risks in large-scale projects.
\end{tcolorbox}

\subsection{Debt: Security Vulnerabilities}

Another concern that is really getting attention in practice is: if we develop a system using a VC approach, how secure will it be? There is plenty of online discussion about security issues via Vibe-coded applications. We also scanned seven early-stage vibe-coded MVPs (developed primarily in Python, JavaScript, TypeScript, and HTML, and ranging from 0.5 to 6 thousand lines of code) using one agentic security scanner \footnote{https://github.com/GPT-Laboratory/Agentic-Code-Security-Auto-Fixer}, which revealed security weaknesses across multiple vibe-coded projects. In total, 970 security issues were detected, including 801 high-severity, 113 medium-severity, and none low-severity. A couple of prominent security issues included path traversal, insecure storage, hard-coded secrets, command injection, cross-site scripting (XSS), insecure deserialization, and missing authentication checks. Overall, unsafe input handling, insecure file operations, and exposed credentials accounted for over 70\% of all findings.


Recent studies support these concerns. Veracode found that nearly half of AI-generated code contained vulnerabilities, with failure rates above 70\% for Java and 38--45\% for Python, C\#, and JavaScript~\cite{techradar2025-veracode}. Trend Micro’s 2025 report further showed that vulnerabilities intensify when AI components are integrated into larger ecosystems, with previously unknown ``zero-day'' exploits appearing in vector databases, inference servers, and container toolkits~\cite{trendmicro2025-security}. These results highlight the importance of embedding continuous, automated security scanning into vibe-coding environments (such as Cursor, Lovable, or Bolt) to detect and remediate vulnerabilities early, promoting secure-by-design practices in AI-assisted development workflows. We observed that some of the VC tools already start embedding the initial version of the security scanner. For instance, Lovable provides a 'Security Scan' feature that identifies security issues and warnings using LLMs and can automatically fix them. One author of this study also highlighted the promising potential of GenAI-based defense mechanisms:

\begin{quote}
{\faQuoteLeft[solid]} \textit{At the same time, GenAI-based evaluation agents can be used as an additional line of defense. These agents can act as an “infinite resource” for scanning applications during development and acceptance phases. White-hat hacker agents simulate ethical security testing by conducting controlled attacks on the system to uncover vulnerabilities early. This helps teams detect insecure patterns before they reach production and keeps regenerated components aligned with security expectations.} {\faQuoteRight[solid]}
\end{quote}

These insights highlight how practitioners increasingly rely on GenAI not only for development acceleration but also as a complementary safeguard. By integrating defensive agents into the workflow, teams can reduce the likelihood that hidden vulnerabilities will escape into production.

\begin{tcolorbox}[
  enhanced,
  breakable,
  colback=white,
  boxrule=0pt,
  left=2mm,
  borderline west={1pt}{0pt}{black!80}
]
\textbf{Takeaway:} Security in vibe coding is a widely shared concern across practice and research. Without explicit safeguards, generated code risks embedding insecure patterns that accumulate into long-term security debt.
Security in vibe coding is a widely shared concern across practice and research. Without explicit safeguards, generated code risks embedding insecure patterns that accumulate into long-term security debt.
\end{tcolorbox}

\subsection{Debt: Maintainability Challenges and Burden}
Maintainability refers to the ease with which software can be understood, corrected, adapted, and evolved over time~\cite{iso25010}. In conventional software engineering, maintainability is supported through modular design, documentation, and architectural consistency ~\cite{ali2018architecture}. In VC, we observed that these qualities are frequently compromised. The produced code appears complete but lacks the structural rationale required for maintainability and evolution. Developers often rely on regenerated outputs rather than internal comprehension, eroding technical ownership and knowledge retention. Over time, redundant files, unused modules, and conflicting logic accumulate, creating a complex and hard-to-navigate codebase. For example, in one of our prototype systems, \textit{a regenerated backend service created a second copy of the \texttt{database.py} module with slightly different function names. Although both versions appeared valid, only one was actually referenced by the API layer. This duplication caused silent runtime failures when the assistant later regenerated related handlers, since it referenced functions from the unused module. Tracing the failure required manual inspection of multiple regenerated files, revealing how quickly structural entropy accumulates when regeneration is not tightly controlled.}

We observed that even modest MVPs that combine front-end logic, backend services, and authentication layers can quickly exceed human comprehensibility when assembled entirely through prompts. Maintainability deteriorates not because of missing functionality but because of \textit{structural entropy} (progressive disorder arising from uncoordinated regeneration) and the accumulation of inconsistencies and artifacts across multiple AI generations. In one observed case, we identified three unused modules and duplicate utility files (e.g., \texttt{utils.py}) that contained overlapping but divergent function names. While targeted use of copilots for isolated features mitigates this risk, fully vibe-coded systems tend to lack cohesion and long-term evolvability.

Once systems reach production, a related issue arises: the \textit{maintenance burden}. AI-generated software often lacks traceability to its originating prompts and model versions. When updates or fixes are required, minor modifications can overwrite prior configurations or reintroduce past errors. Regenerated code fragments may also diverge from earlier data structures or APIs, breaking integrations. This phenomenon is frequently described as \textit{AI-generated technical debt}, reflecting the cost of sustaining a codebase that evolves faster than it can be understood or documented~\cite{mahendra2025-vibecoding}. One author of this study also summarized his experience about maintainability and possible solutions for it:  

\begin{quote}
{\faQuoteLeft[solid]} \textit{In our own development experience, we have seen maintainability burden grow quickly under VC-driven workflows, as rapid regeneration cycles can exceed the capacity of teams to track and coordinate changes. To manage this, we found value in using refactoring and migration agents that automate cleanup and restructuring tasks, reinvesting the operational gains from fast development into these maintenance tools. We also observed that CI/CD pipelines are becoming more proactive, with continuous analysis and scanning tools that detect issues before they spread. This direction is likely to become increasingly important as systems scale.} {\faQuoteRight[solid]}
\end{quote}

As GenAI workflows continue to scale, sustaining software quality will increasingly depend on the tight integration of automated maintenance, scanning, and regeneration control mechanisms.

\begin{tcolorbox}[
  enhanced,
  breakable,
  colback=white,
  boxrule=0pt,
  left=2mm,
  borderline west={1pt}{0pt}{black!80}
]
\textbf{Takeaway:}: 
Growing industry experience suggests that automated refactoring agents and proactive CI/CD analysis will play an increasingly important role in managing this maintainability burden at scale.
\end{tcolorbox}

\subsection{Debt: Testing Gaps}
VC platforms such as Cursor can generate unit, integration, and end-to-end tests when prompted, creating the impression of automated validation. In practice, these tests often cover only “happy-path” cases and omit edge conditions, failure states, and error-handling scenarios. Some outputs are merely trivial stubs or mocks that run without verifying meaningful functionality, creating a false sense of completeness.

We experienced that test quality varies greatly depending on prompt phrasing and model version. Generated tests often duplicate or contradict one another after multiple regeneration cycles, particularly when the underlying code evolves faster than the tests themselves.
This instability weakens regression coverage and makes continuous integration unreliable. In one of our prototype systems, the VC tools assistant-generated test suite for an authentication workflow passed consistently, even though the login was broken in practice. The tests relied entirely on mocked responses created by the model itself rather than exercising the real session logic. As a result, invalid tokens were accepted, and expired sessions were never rejected. This mismatch went unnoticed until manual exploratory testing revealed that authenticated pages could not be accessed at all.

Most current VC tools lack systematic traceability between requirements and test cases. Generated tests usually focus on functional correctness, while non-functional attributes such as performance, security, and scalability are often left untested. This gap becomes especially serious in multi-service systems, where interface mismatches and timing issues can remain hidden until deployment~\cite{microsoft2024-ai-testing, ieee2024-aiqase}. However, some newer vibe coding tools have begun adding early forms of non-functional requirement agents that scan specifications at different stages of the SDLC, such as GDPR checkers, compliance analyzers, and usability evaluators. As these capabilities mature, automated NFR analysis is likely to become a standard feature in future VC environments.

We have experienced in practice that maintaining meaningful test coverage in VC requires human oversight, explicit test-to-requirement mapping, and revalidation after each major regeneration cycle. A practical approach is to combine three layers of testing discipline: (1) AI-generated unit tests complemented by human-written edge cases; (2) contract and interface tests across services to ensure stable integration; and (3) continuous performance and security gates embedded within CI pipelines. Without such multi-layer safeguards, auto-generated test suites provide an illusion of reliability while silently allowing errors to propagate into production.

\begin{tcolorbox}[
  enhanced,
  breakable,
  colback=white,
  boxrule=0pt,
  left=2mm,
  borderline west={1pt}{0pt}{black!80}
]
\textbf{Takeaway:}  Vibe coding can produce test suites automatically, but they are often unstable, incomplete, and misaligned with evolving codebases. Effective validation requires human-guided test planning, cross-checking against requirements, and automated regression testing governance to ensure testing evolves alongside the system.
\end{tcolorbox}

\subsection{Debt: Deployment Fragility (CI/CD and Operations)}
Fast, prompt-to-code workflows often outpace the guardrails that make deployment safe. We observed that code bases frequently reach CI/CD pipelines in unstable states. As a result, builds, packaging, and environment assumptions often break at integration time. Industry reports (e.g., ~\cite{infosprint2025-cicd, microsoft2024-devops}) warn that this unpredictability amplifies release risk, operational toil, and long-term debt unless CI/CD, automated testing, and observability are embedded early in the vibe coding workflow. 

In our experience, we also observed that these challenges became tangible when sub-teams generated their own services through iterative prompting. The developed services worked in isolation but failed during integration. We experienced container builds breaking due to mismatched dependencies, inconsistent environment variables, and overlapping network ports. Based on our experience, we recommend treating AI-generated code as a first draft that must pass strict automated gates such as linting, type checks, and tests before merging. Some teams also employ AI reviewers within CI pipelines to analyze pull requests and enforce quality ~\cite{flatt2025-cicd}. CI pipelines themselves should be hardened with secure runners, isolated secrets, and audit trails to prevent leaks and tampering~\cite{futurecode2025risks}. 
We noted that operational fragility extends post-deployment. For example, vibe-coded systems often include undeclared dependencies or inconsistent runtime configurations that fail under load. One possible strategy to address this is to embed observability through structured logging, health probes, dashboards, and automated rollback mechanisms to detect and recover from failures early~\cite{infosprint2025-cicd}.

\begin{tcolorbox}[
  enhanced,
  breakable,
  colback=white,
  boxrule=0pt,
  left=2mm,
  borderline west={1pt}{0pt}{black!80}
]
\textbf{Takeaway:} Deployment fragility arises when rapid AI-generated outputs meet real-world integration and operations. Sustainable scaling requires secure, automated CI/CD pipelines, strong observability, and human oversight to manage inconsistencies and prevent accumulating technical debt.
\end{tcolorbox}

\begin{figure*}[t]
\centering
\begin{ScenarioBox}[Industry Perspective: When Vibe Coding Helps – and When It Hurts]
\small
\begin{itemize}
  \item VC works well for internal tools, prototypes, and experiments, but our industry teams treat it as unsafe by default for customer-facing systems without extra hardening and review.
  \item Regeneration speed routinely outpaces human review: a single “fix this” prompt can rewrite large parts of the codebase before architects or testers have seen the previous version.
  \item Each regeneration round risks eroding architecture boundaries, duplicating logic, and breaking service contracts, so integration pain grows much faster than the visible feature set.
  \item Early VC-based MVPs that “look done” on the surface have repeatedly failed basic security scans, exposing issues such as hard-coded secrets, missing authentication, and unsafe input handling.
\end{itemize}
\end{ScenarioBox}
\label{fig:refactoring-iteration-overload}
\end{figure*}

\begin{table*}[!htb]
\scriptsize
\centering
\caption{Guideline Matrix for Sustainable Vibe Coding: Issues, Causes, and Role-Specific Countermeasures ( 
\faUsers\ = guidance primarily for engineering teams; 
\faTools\ = guidance primarily for tool/platform builders; 
\faUserTie\ = guidance primarily for CIO/CTO and organizational leadership.)}
\begin{tabular}{p{2cm} p{2.6cm} p{3cm} p{3cm} p{6.4cm}}
\toprule
\textbf{Dimension} & \textbf{Perceived value} & \textbf{Debt} & \textbf{Potential resolution} & \textbf{Guideline / Recommendation} \\
\midrule

\multirow{4}{*}{
\centering\shortstack[c]{%
\textbf{Model}\\[2pt]
\includegraphics[height=0.5cm]{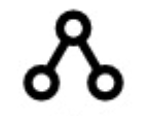}
}%
}
& Fast ideation & Requirements ambiguity & Alignment tuning & \faUsers\ Use constrained prompt templates; validate intent examples; explicitly capture ASRs/NFRs as quality‐attribute scenarios linked to tests and architecture. \\
& Creative expansion & Hallucinated or speculative logic & Safe generation policies & \faTools\ Bound generation scope via explicit constraints (security, performance, architecture); discourage speculative abstractions that inflate rework. \\
& Plausible outputs & Misaligned reasoning & Explainability tools & \faTools\ Apply attribution/visualization to inspect model decisions; gate merges when explanations diverge from intended design or requirements. \\
& Scope expansion excitement 
& Overextended feature scope and distraction from core requirements
& Value slicing / scope triage
& \faUsers\ Triaging AI-proposed features into must‐have vs. optional; pruning speculative additions that lack stakeholder validation or architectural grounding. \\
\multicolumn{5}{p{18.0cm}}{\textit{Model level captures how acceleration in idea realization and creativity can lead to ambiguity or misalignment. Constraint, validation, and explainability sustain semantic stability.}} \\

\midrule

\multirow{5}{*}{
\centering\shortstack[c]{%
\textbf{Process}\\[2pt]
\includegraphics[height=0.5cm]{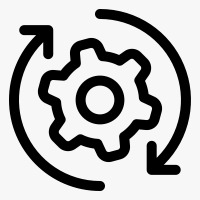}
}%
}
& Rapid iteration & Architectural inconsistency & Prompt versioning / Continuous validation & \faUsers\ Store prompts with code; link requirements $\leftrightarrow$ tests $\leftrightarrow$ architecture; enforce regression and design reviews on regenerations. \\
& Auto testing & Testing gaps & Continuous validation & \faUsers\ Combine AI‐generated and human‐written tests; ensure requirement–test traceability; automate regression validation in CI pipelines. \\
& Fast delivery & Deployment fragility (CI/CD) & CI/CD hardening & \faUsers\ Enforce reproducible builds, pinned dependencies, and policy‐as‐code gates; embed observability, rollback plans, and secure secret management. \\
& Unified conversational flow 
& Blurred boundaries between requirements, pseudocode, design, and tests
&Scenario alignment (BDD/ATDD)
& \faUsers\ Update BDD/ATDD scenarios whenever requirements or regenerated components change; treat scenarios as canonical specifications for validation. \\
&Visually polished MVPs that appear production‐ready
&Illusion of completeness with missing tests, documentation, and architectural rationale 
& Release‐readiness checklists
& \faUserTie\ Use Definition‐of‐Done including documents, tests, architectural sketches, and ops checks before marking any MVP “production‐ready.” \\
\multicolumn{5}{p{18.0cm}}{\textit{Process level shows how iteration speed and automation increase productivity but weaken integration and testing. Versioning, validation, and hardened pipelines sustain process reliability.}} \\

\midrule

\multirow{4}{*}{
\centering\shortstack[c]{%
\textbf{Platform}\\[2pt]
\includegraphics[height=0.5cm]{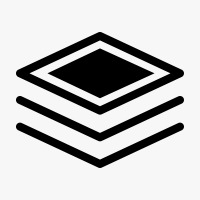}
}%
}
& Easy integration & Security vulnerabilities & Integrated QA pipelines & \faTools\ Run SAST/DAST/SBOM scans automatically; block risky imports or open configs; embed automated security gates in workflows. \\
& Quick reuse & Security / provenance debt & Provenance logging & \faTools\ Capture prompt hash, model/version, parameters, time, and context for each artifact to support auditability and rollback. \\
& Fast regeneration & Integration instability & Semantic differencing & \faTools\ Compare regenerated code semantically; highlight intent‐level changes rather than line diffs to prevent drift and merge conflicts. \\
& GenAI-based evaluation agents for continuous scanning
& Overreliance on automated checks and blind spots in agent coverage
& Hybrid human–AI security review
& \faTools\ Use evaluation agents as supplementary defence; periodically validate agent findings with manual penetration tests and adjust scanning policies.\\
\multicolumn{5}{p{18.0cm}}{\textit{Platform level highlights how seamless reuse and integration accelerate work but introduce security and provenance debt. Logging and semantic differencing maintain control and trust.}} \\

\midrule

\multirow{4}{*}{
\centering\shortstack[c]{%
\textbf{Organisation}\\[2pt]
\includegraphics[height=0.5cm]{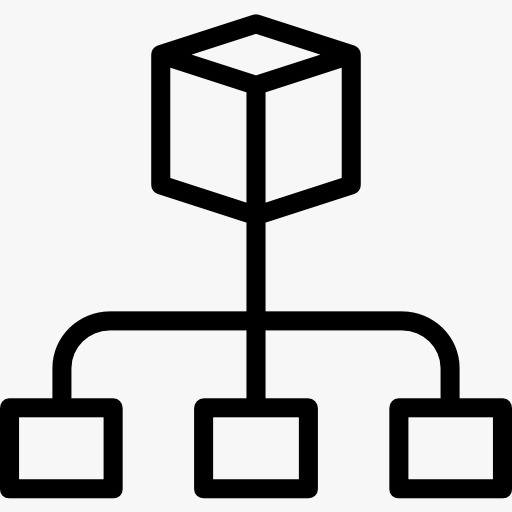}
}%
}
& Fast onboarding & Maintainability burden & Skill continuity / Human‐in‐the‐loop & \faUserTie\ Train developers in AI literacy, prompt craft, and SE fundamentals; require comprehension checks and peer reviews; preserve rationale and documentation. \\
& Easy delegation & Knowledge erosion & Accountability policies & \faUserTie\ Define roles, approval gates, and audit trails; assign ownership for AI contributions to sustain accountability and prevent knowledge loss. \\
& Prompt‐based participation and inclusive team engagement
& Role confusion and erosion of engineering judgment 
& Participation guardrails
& \faUserTie\ Define approval permissions for AI-generated changes; ensure pairing with senior reviewers; enforce sign-off for high-impact code generations. \\
& Rapid regeneration flexibility
& Structural entropy and weakened technical ownership
&Refactoring/migration agents + traceability
& \faUsers\ Schedule AI-assisted maintenance sprints; log prompts and model versions; reconcile duplicates and remove unused modules. \\
\multicolumn{5}{p{18.0cm}}{\textit{Organisation level shows how automation accelerates productivity but risks skill decay and ownership loss. Training and human oversight sustain maintainability.}} \\

\midrule

\multirow{4}{*}{
\centering\shortstack[c]{%
\textbf{Environment}\\[2pt]
\includegraphics[height=0.5cm]{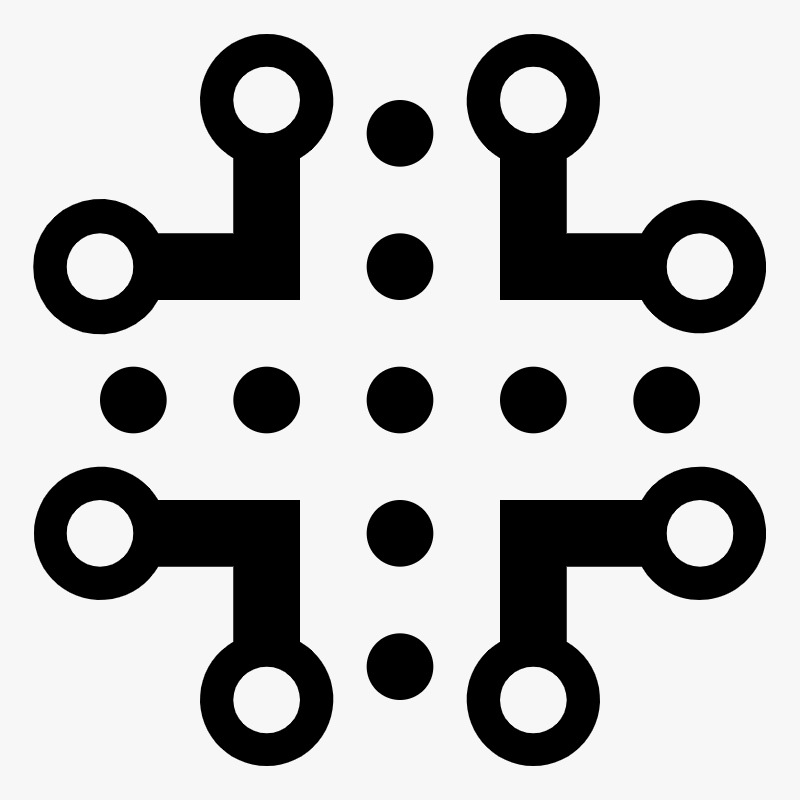}
}%
}
& Fast experimentation & Compliance / sustainability debt & Compliance frameworks & \faUserTie\ Align with emerging AI/SE standards; maintain audit evidence (provenance, approvals, SBOM); ensure environmental and regulatory adherence. \\
& Quick portability & Platform portability debt & Platform abstraction & \faTools\ Encapsulate provider APIs; use adapters/contracts to reduce lock‐in and portability risk across infrastructures. \\
& Easy data reuse & Data/IP governance debt & Provenance control & \faUserTie\ Track data/IP lineage, licenses, and reuse; monitor energy and maintainability metrics for long‐term sustainability. \\
& Early NFR/compliance scanning agents
& Superficial or incomplete compliance assessment 
& Governance of AI-based compliance checks
& \faUserTie\ Treat compliance agents as advisory; map outputs to standards (GDPR, ISO, safety); maintain human audits and compliance evidence. \\
\multicolumn{5}{p{18.0cm}}{\textit{Environment level captures how fast experimentation and reuse improve agility but raise compliance and governance risks. Abstraction and provenance control maintain long-term viability.}} \\

\bottomrule
\end{tabular}
\label{tab:flow-debt-matrix}
\end{table*}

\section{Causes and Guidelines}
\label{sec:causes-and-countermeasures}
This section summarises the key issues observed in vibe-coded systems, their causes, and guidelines to address them. These insights emerged directly from our experience building and evaluating VC-assisted developed systems. Table~\ref{tab:flow-debt-matrix} presents these elements across five dimensions: \textit{model, process, platform, organisation,} and \textit{environment}. The left-hand columns describe why failures emerge, while the right-hand columns outline corresponding countermeasures and practical guidelines.

At the model level, for example, slight variations in phrasing or context routinely produced divergent implementations. Structured prompt constraints and alignment tuning reduced this volatility by directing the model toward consistent architectural and behavioral interpretations. Process-level inconsistencies were best addressed by maintaining explicit traceability between prompts, requirements, tests, and regenerated code, combined with continuous validation. Platform fragility was mitigated through provenance logging, semantic differencing, and integrated security scanning. Organizational challenges, such as knowledge erosion or weakened technical ownership, were reduced through human oversight, clear approval policies, and hybrid skill development. Environmental risks, including compliance and data-governance gaps, required explicit documentation and auditability of AI-generated artifacts.

Taken together, these observations reinforce that sustainable VC requires more than incremental model improvement. It depends equally on disciplined engineering processes, tool chains, transparent collaboration, and responsible governance. Table~\ref{tab:flow-debt-matrix} also synthesizes these perspectives into a unified guideline matrix that practitioners and tool builders can apply directly in their workflows.

\subsection*{Sustainable Vibe Coding Checklist}
To complement the guideline matrix in Table~\ref{tab:flow-debt-matrix}, we propose a \textit{sustainable vibe coding checklist}. This checklist introduces additional recommendations that support the matrix and align with standard software engineering practices. It synthesizes the most actionable steps across the five dimensions of the matrix and serves as a quick reference for teams adopting VC workflows.


\begin{itemize}
    \item[\large$\square$] \textbf{Define project boundaries }and remove AI-suggested features that stakeholders have not approved.
    \item[\large$\square$] \textbf{Write clear and structured prompts} that capture functional and non-functional needs.
    \item[\large$\square$] \textbf{Maintain traceability} by linking prompts, generated code, and the model settings used.
    \item[\large$\square$] \textbf{Update BDD or acceptance scenarios} whenever requirements or regenerated parts change.
    \item[\large$\square$] \textbf{Run regression checks} and compare results before accepting regenerated code.
    \item[\large$\square$] \textbf{Strengthen pipelines} by ensuring reproducible builds, fixed dependencies, and easy rollback steps.
    \item[\large$\square$] \textbf{Monitor security} using regular automated scans, with AI-based checks used only as supportive tools.
    \item[\large$\square$] \textbf{Request senior} review for high-impact regenerated components and clarify ownership roles.
    \item[\large$\square$] \textbf{Track data and IP} sources and keep records needed for compliance.
    \item[\large$\square$] \textbf{Check} maintainability, energy use, and long-term sustainability during routine operations.
\end{itemize}

\section{Future Outlook and Conclusion}

VC is progressing through clear capability milestones, shifting from basic code suggestions to whole application generation by 2025. New tools now build front-end and back-end systems from natural-language instructions and increasingly support architectural and deployment choices. These advances also bring risks, as features that boost developer flow can introduce instability and technical debt. The flow–debt balance reflects this tradeoff, but improvements in context handling, retrieval, interpretability etc. are gradually reducing these issues.

\vspace{0.5 em}
\textbf{Vision for future research on VC}: Future research on vibe coding should move from general impressions to evidence-based understanding.

  \begin{itemize}
      \item Analyzing vibe-coded repositories on platforms such as GitHub, GitLab, and Stack Overflow to identify vulnerabilities, code smells, and quality trends.
      \item Surveying vibe-coded practitioners to understand benefits, trust levels, effort, and practical challenges.
      \item Collaborating on vibe-coded research between academia and industry to build shared datasets, run joint CI/CD experiments, and support responsible adoption.
  \end{itemize}

An additional long-term direction involves exploring GenAI-based operating environments, where the entire software stack functions as a conversational VC platform, enabling continuous regeneration but also raising new challenges for stability, governance, and maintainability.

Future research and development incorporating empirical evidence, metrics, and case-study insights can guide trustworthy adoption of sustainable vibe-coding practices.

\section*{Acknowledgment of AI-Assisted Tools}
The authors acknowledge that large language models, including ChatGPT and similar AI-based assistants, were used solely for grammar refinement, phrasing improvement, and proofreading during the preparation of this manuscript. Grammarly was additionally used for grammar and style checking. No AI system was used to generate technical content, figures, images, data, analysis, or code appearing in this article. All scientific claims, interpretations, and conclusions are the authors' own.

\bibliographystyle{IEEEtran}
\bibliography{references}

\end{document}